\documentclass[aps,12pt,showpacs,twoside]{revtex4}

\usepackage[english]{babel}
\usepackage{graphicx}
\usepackage{amsmath,amsfonts,amssymb,amsthm}
\usepackage{enumerate}
\usepackage{epsfig}

\parskip=\medskipamount

\newcommand{\eq}[1]{(\ref{#1})}
\newcommand{\fig}[1]{Fig. \ref{#1}}

\newcommand{\be}{\begin{equation}}
\newcommand{\ee}{\end{equation}}

\newcommand\disp{\displaystyle}

\newcommand{\la}{\left<}
\newcommand{\ra}{\right>}

\newcommand{\eps}{\varepsilon}
\newcommand{\N}{\parallel}
\newcommand{\T}{\perp}

\def\runninghead#1#2{\pagestyle{myheadings}
\markboth{{\protect\sl{\quad #1}}\hfill} {\hfill{\protect\sl{#2\quad}}}}

\begin{document}

\runninghead{\sl A. Gorsky, S. Nechaev, V. Poghosyan, V. Priezzhev}{From elongated spanning trees
to vicious random walks}

\begin{flushright}
ITEP--TH--22/12
\end{flushright}

\title{From elongated spanning trees to vicious random walks}
\author{A. Gorsky$^1$}

\author{S. Nechaev$^{2,3}$}

\author{V.S. Poghosyan$^4$}

\author{V. B. Priezzhev$^5$}

\affiliation{$^1$ITEP, B. Cheryomushkinskaya 25, 117218 Moscow, Russia \\
$^2$LPTMS, Universit\'e Paris Sud, 91405 Orsay Cedex, France \\
$^3$P.N. Lebedev Physical Institute of the Russian Academy of Sciences, 119991 Moscow, Russia \\
$^4$Institute for Informatics and Automation Problems NAS of Armenia, 375044 Yerevan, Armenia \\
$^5$ Bogolubov Laboratory of Theoretical Physics, Joint Institute for Nuclear Research, 141980
Dubna, Russia}

\begin{abstract}

Given a spanning forest on a large square lattice, we consider by combinatorial methods a
correlation function of $k$ paths ($k$ is odd) along branches of trees or, equivalently, $k$
loop--erased random walks. Starting and ending points of the paths are grouped in a fashion a
$k$--leg watermelon. For large distance $r$ between groups of starting and ending points, the ratio
of the number of watermelon configurations to the total number of spanning trees behaves as
$r^{-\nu} \log r$ with $\nu = (k^2-1)/2$. Considering the spanning forest stretched along the
meridian of this watermelon, we see that the two--dimensional $k$--leg loop--erased watermelon
exponent $\nu$ is converting into the scaling exponent for the reunion probability (at a given
point) of $k$ (1+1)--dimensional vicious walkers, $\tilde{\nu} = k^2/2$. Also, we express the
conjectures about the possible relation to integrable systems.

\end{abstract}


\maketitle

{\small \tableofcontents}

\section{Introduction}

\begin{figure}[ht]
\epsfig{file=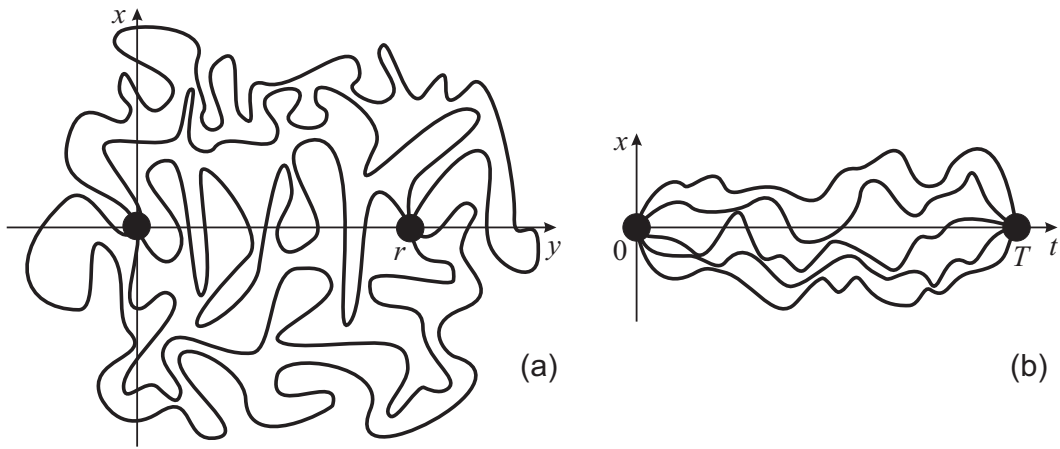, width=14cm}
\caption{a) Symmetric $k$--leg watermelon in 2D made of loop--erased random walks; b) Stretched
watermelon (``zucchini'') of $k$ directed (1+1)D vicious walkers.}
\label{fig:0}
\end{figure}

Enumeration of spanning trees on $d$--dimensional grids is a classical problem of the combinatorial
graph theory \cite{Kirhhoff}. According to the Kirchhoff theorem, the number of spanning tree
subgraphs on a lattice is given by minors of the discrete Laplacian matrix $\Delta$ of this
lattice. Considering a modified matrix $\Delta'$, which differs from $\Delta$ by a finite number of
elements, one can define local correlation functions of spanning trees in any dimension. Besides of
the combinatorial methods, in the two--dimensional case the conformal invariance can be used, which
allows to evaluate the nonlocal spanning tree correlations of special topology. For instance, using
a mapping of the $q$--state Potts model to the Coulomb gas, and taking the limit $q \to 0$, Saleur
and Duplantier \cite{SalDup} have found the correlation function $W_k(r)$ of a $k$--path
``connectivity'' in the spanning tree. The $k$--path connectivity means that, given two lattice
points $A$ and $B$ separated by the distance $r$ ($r\gg 1$), one can find $k$ paths along branches
of the spanning tree from a vicinity of the point $A$ to that of the point $B$. Such a topology is
called in physical literature the $k$--leg watermelon and each path along the branches is a
loop--erased random walk (LERW). Schematically, the $k$--leg watermelon is depicted in the figure
\fig{fig:0}a. In this work we will be mostly interested in the asymptotic behavior of $W_k(r)$ for
large separations $r\gg 1$.

In the frameworks of the conformal field theory, the correlation function of $k$ connected clusters
tied together at the extremities separated by the distance $r$, has the following scaling behavior:
\be
W_k(r) \sim r^{-2x_{k}}
\label{eq:2}
\ee
where $x_k$ is the conformal dimension of the scalar primary fields, and $k$ is the number of
clusters. By mapping to the $q$--state Potts model with arbitrary $q$, one allows to see that each
cluster of the Potts model in the limit $q \to 0$ contains a path from one extremity to another.
The values of scaling dimensions, $x_{p,q}$, have been computed as
\be
x_{p,q} = 2 h_{p,q}; \quad h_{p,q}=\frac{\left[(m+1)p-mq\right]^2-1}{4m(m+1)}
\label{eq:3}
\ee
where $h_{p,q}$ are the weights belonging to the Kac table and $p,q$ are integers in the minimal
block, $1\le p\le m-1$. The parameter $m$ in \eq{eq:3} is fixed by the central charge $c$,
\be
c=1-\frac{6}{m(m+1)}
\label{eq:4}
\ee
For $c=-2$ (i.e., $m=1$) one has
\be
h_{p,q} =\frac{(2p-q)^2-1}{8}
\label{eq:5}
\ee
and for $(p,q)=(k/2,0)$ the large--distance scaling behavior of the $k$--path correlation function
is
\be
W_k(r) \sim r^{-4h_{k/2,0}} = r^{-\frac{k^2-1}{2}}
\label{eq:6}
\ee

A combinatorial derivation of the exponent $\nu = (k^2-1)/2$ for the $k$--leg watermelon remains
still an open problem. However, a ``minor'' modification of the problem makes it solvable. Instead
of the $k$--leg watermelon embedded into the {\em single--component} spanning tree, one may
consider the {\em two--component} spanning tree and put the watermelon into one of these
components. Then, for sufficiently large components, the partition function of $k$--leg watermelon
configurations for odd $k$ is given by the determinant of $k \times k$ matrix, whose entries are
the Green functions $\Delta^{-1}$ of the discrete Laplacian $\Delta$.

The partition function of the 3--leg watermelon on the two--component spanning tree has been
computed for the first time in \cite{Pri94} where the 3--leg watermelon, called the
$\theta$--graph, has been used for exact determination of the height probabilities in the Abelian
sandpile model \cite{Bak,Dhar}. The $k$--leg generalization of the $\theta$--graph was the subject
of the work \cite{Iva}, where the authors have derived a determinant expression for the $k$--leg
watermelon also embedded into one component of the two--component spanning tree. Calculating
$W_k(r)$ for few small values of $k$ ($k=1,3,5$), and guessing the structure of the series
expansion of $W_k(r)$ at $r\gg 1$, they have suggested the generic expression
\be
W_k(r) \sim r^{-\frac{k^2 -1}{2}}\ln r
\label{wkrlog}
\ee
where the logarithmic factor comes from the two--component topology of the problem (see e.g.
\cite{KtitIvaPri}). An exact derivation of \eq{wkrlog} needs more careful analysis of the resulting
determinant. In our work we show how the asymptotic correlation function \eq{wkrlog} can be derived
exactly from the corresponding determinant expression.

The determinant expression of the $k$--leg watermelon in the two--component spanning tree can be
generalized to the situation when the spanning trees are elongated in some direction. Ascribing
additional weights to the lattice bonds along some direction, one can obtain an asymmetric
Laplacian $\Delta^{(\eps)}$ which depends on the degree of elongation $\varepsilon$ ($0 \leq \eps
\leq 1/4$), where $\eps = 0$ corresponds to the isotropic lattice and $\eps = 1/4$ -- to the fully
elongated case. The Kirchhoff theorem, being applied to $\Delta^{(\eps)}$, gives the partition
function of {\em anisotropic spanning trees}.

The anisotropy affects the statistical properties of branches of the trees and, therefore, the
loop--erased random walks constructed as paths on the elongated branches. At extreme stretching
$\eps = 1/4$ along a diagonal of the square lattice, the loop--erased random walk becomes obviously
the simple Bernoulli process with equal probabilities to take one of two possible directions. We
consider the $k$--leg watermelon statistics for $\eps$ between 0 and $1/4$. In the limiting case
$\eps = 1/4$, one may expect that the loop--erased random walks constituting the watermelon behave
like vicious walkers \cite{Fisher} and the watermelon geometry corresponds to the reunion at a
given point in the space of $k$ walkers after equal number of steps made by all walkers. The
elongated $k$--leg watermelon is drawn in the figure \fig{fig:0}b. In contrast to non-stretched
watermelon, it is natural to call it ``zucchini''. According to \cite{Fisher} and
\cite{HuseFisher}, the correlation function $W_k(r)$ at $r \gg 1$ is
\be
W_k(r) \sim r^{-\frac{k^2}{2}}
\label{wkrFisher}
\ee

If the behavior of the watermelon at extreme elongation is rather clear, the asymptotics of
$W_k(r)$ at small elongations is less obvious. The presence of the bias, produced by an external
field, forces the loop--erased walk to follow the direction of the field, thus breaking the
conformal invariance and making the two--dimensional system effectively (1+1)--dimensional. At
small $\eps$, the external field is a perturbation of the logarithmic conformal field theory with
$c=-2$, and determination of the exponent in \eq{wkrFisher} by the CFT methods becomes a
non--trivial problem. So, the main goal of our work consists in examination how the watermelon
exponent is converting into the scaling exponent for the reunion probability of $k$ vicious walkers
in the interval $0\le \eps \le 1/4$.

Another intriguing application of our work concerns the ``interpolation between CFT and RMT''. It
is known that the isotropic LERW is described by the 2D $SLE_k$ with $k=2$ \cite{Lawler}, while the
directed mutually avoiding (``vicious'') walks deal with statistics described by the random matrix
ensembles. Changing the degree of extension, $\eps$, we pass from the $c=-2$ CFT to the RMT
ensemble.

\section{Watermelons of loop--erased random walks (LERW), spanning trees and Wilson algorithm}

In order to make the contents of the paper as self--contained as possible, we outline in this
section some connections between spanning trees and the loop--erased random walks.

Consider a finite square lattice $\mathcal{L}$ with vertex set $V$ and edge set $E$. Select a
tagged subset $W$ ($W \subset V$), which is called ``the set of dissipative vertices''. Let
$\mathcal{P}=[u_0, u_1, u_2, \ldots, u_n]$ be a collection of vertices (a trajectory) passed by a
$n$--step random walk on $\mathcal{L}$, started from $u_0$ and ended upon hitting $u_n \in W$. The
loop--erasure $\mathrm{LE}(\mathcal{P}) = [\gamma_0, \gamma_1, \gamma_2, \ldots, \gamma_m]$ of
$\mathcal{P}$ is defined by removing loops from $\mathcal{P}$ in a chronological order. Note that
the order in which we remove loops is important, it is uniquely fixed by the condition that the
loop is removed immediately as it is created when we follow the trajectory $\mathcal{P}$. The
ordered collection of vertices $\mathrm{LE}(\mathcal{P})$ is called ``the loop--erased random
walk'' (LERW) on $\mathcal{L}$ with dissipation $W$.

Wilson \cite{Wilson} has proposed an algorithm to generate the ensemble of uniformly distributed
spanning trees by LERWs. It turns out to be useful not only as a simulation tool, but also for
theoretical analysis. The algorithm runs as follows. Pick up an arbitrary ordering $V \setminus W =
\{v_1, \ldots, v_N \}$ for the vertices in $\mathcal{L}$. Inductively, for $i = 1,2,\ldots, N$
define a graph $S_i$ ($S_0=W$) which is a union of $S_{i-1}$ and a (conditionally independent) LERW
from $v_i$ upon hitting $S_{i-1}$. If $v_i \in S_{i-1}$, then $S_i=S_{i-1}$. Regardless of the
chosen order of the vertices, $S_N$ is a sample of uniformly distributed random spanning forests on
$\mathcal{L}$ with the set of roots $W$. A spanning forest with one root is a spanning tree. A
spanning forest with a fixed set of roots can be considered as a spanning tree, if we add an
auxiliary vertex and join it to all the roots. When the size of the lattice tends to infinity, the
boundary effects vanish, so we can neglect the details of the boundary. {\em We do not distinguish
between spanning forests and spanning trees in this case, assuming all boundary vertices are
connected to an auxiliary vertex.}

Consider the set of $k$ bulk vertices $I_k=\{i_1,i_2,\ldots, i_k\}$. The Wilson algorithm on
$\mathcal{L}$ with the dissipation $W = \partial\mathcal{L} \cup I_k$ generates $k+1$--component
spanning forest with roots $i_1,i_2,\ldots, i_k$ and an auxiliary vertex connected to the boundary
vertices. Given $k$, and introducing another set of $k$ vertices $J_k=\{j_1,j_2,\ldots, j_k\}$, we
can enumerate all spanning forests, for which the vertices in $J_k$ belong to the different
components rooted at vertices in $I_k$.

Construct now an auxiliary Laplacian on $\mathcal{L}$ using the bridge trick (\cite{Pri94}).
Specifically, define the unperturbed Laplacian of the square lattice, $\Delta$, and the perturbed
Laplacian, $\Delta'$, which differs from $\Delta$ by ``defects'' (bridges) with the weights
$-\eta$,
\be
\Delta_{ij} =
\begin{cases}
4 & \mbox{if $i=j$} \\ -1 & \mbox{if $i,j$ are nearest neighbors} \\ 0 & \mbox{otherwise}
\end{cases}; \;
\Delta_{ij}' =
\begin{cases}
4 & \mbox{if $i=j$} \\ -1 & \mbox{if $i,j$ are nearest neighbors} \\ -\eta & \mbox{if $i,j$
is a pair $(i_s,j_s)$, $s=1,...,k$ } \\ 0 & \mbox{otherwise}
\end{cases}
\label{dd}
\ee
One can check that the combination
\be
\lim_{\eta \to \infty} \frac{1}{\eta^k} \det \Delta'
\ee
gives the sum over $k+1$--component spanning forests $T$ having $k$ bulk roots in the lattice
points $I_k=\{i_1,i_2,\ldots, i_k\}$. Each forest $T$ enters into the sum with the sign
$(-1)^{c(T)}$, where $c(T)$ is the number of cycles which appear in $T$ due to adding bridges
$(i_s,j_s)$, $s=1,2,\ldots, k$ -- see \cite{Pri94}.

\begin{figure}[ht]
\epsfig{file=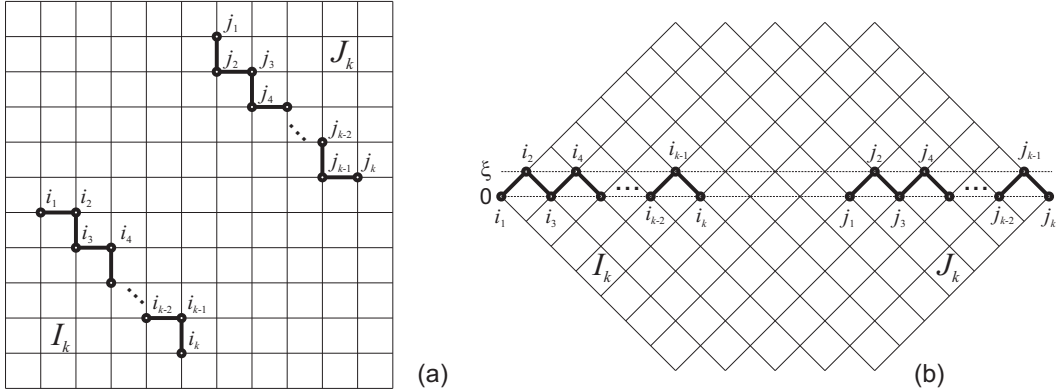,width=14cm}
\caption{The $k$ nonintersecting loop--erased random walks begin in the set $I_k$ and end in the
set $J_k$: (a) Sets $I_k$ and $J_k$ ($k$ is odd) chosen as parallel zigzags on the square lattice;
(b) Sequential zigzags $I_k$ and $J_k$ oriented along the horizontal axis.}
\label{fig:1}
\end{figure}

\begin{figure}[ht]
\epsfig{file=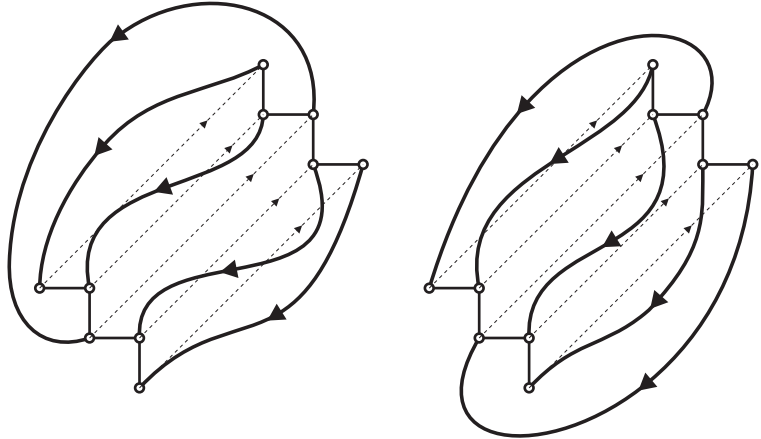,width=10cm}
\caption{Possible closures of zigzags by loop--erased random walks.}
\label{fig:2}
\end{figure}

Following Ivashkevich and Hu \cite{Iva}, we chose sets $I_k$ and $J_k$ as zigzags (or ``fences'')
with odd $k$ -- see \fig{fig:1}a. Then, all possible  configurations of cycles on the bridges
contain either 1 or $k$ cycles -- see \fig{fig:2} and, therefore, $(-1)^{c(T)} = -1$. For fixed $k$
and the large distance between sets $I_k$ and $J_k$, the bunch of lattice paths from $J_k$ to $I_k$
has a form of a watermelon. Connecting the neighboring points in the set $I_k$ (and in the set
$J_k$), we obtain a two--component spanning tree with the watermelon embedded into one component.
Thus, we have
\be
-\lim_{\eta \to \infty} \frac{1}{\eta^k} \det \Delta' = \mbox{number of $k$--leg
watermelons}
\label{epsilon}
\ee

The matrix $\Delta'$ can be represented as $\Delta'=\Delta + B$, where the matrix $B$ has nonzero
elements $-\eta$ for pairs $(i_s,j_s)$, $s=1,2,\ldots,k$. To evaluate the determinant $\Delta'$, we
use the known formula \cite{MajDhar}
\be
\frac{\det \Delta'}{\det \Delta} =\det(\mathbb{I} + B G),
\label{defect_delta}
\ee
where $\mathbb{I}$ is the unit matrix, $G$ is the Green function $G=\Delta^{-1}$ and the nonzero
part of the matrix $B$ is the $k \times k$--matrix. Inserting \eq{defect_delta} into \eq{epsilon},
we obtain the ratio between the number of two--component spanning trees containing the $k$--leg
watermelon and the total number of spanning trees on lattice $\mathcal{L}$:
\be
W(I_k,J_k)= \det G(I_k,J_k)
\label{basic}
\ee
where rows and columns of matrix $G(I_k,J_k)$ are labelled by indices  $i_1,i_2,\ldots, i_k$ and
$j_1,j_2,\ldots, j_k$.

\section{Isotropic LERW watermelon}

The correlation function $W(I_k,J_k)$ at large separation, $r$, between the sets $I_k$ and $J_k$
($k$ is odd) takes an asymptotic form $W_k(r) \sim r^{-\nu}\ln r$ \cite{Iva}, where $\nu$ is a
universal lattice--independent exponent. To find $\nu$, we have to specify the sets $I_k$ and $J_k$
on the square lattice and consider then the continuous limit $r \rightarrow \infty$. Following
\eq{basic} we should define the matrix $G(I_k,J_k)$.

The Green function $G_{x,y}$  on a square lattice is the solution of the equation
\be
-\frac{1}{4}\Big(G_{x-1,y}+G_{x,y-1}+G_{x+1,y}+G_{x,y+1}\Big)+G_{x,y}=\delta_{x,0}\delta_{y,0}
\label{laplacian2}
\ee
The integral form of $G_{x,y}$ is
\be
G_{x,y} = \frac{1}{8\pi^2}\int_{-\pi}^{\pi} d\alpha \int_{-\pi}^{\pi} d\beta \; \frac{ \cos(\alpha
x) \cos(\beta y)}{2-(\cos\alpha + \cos\beta)},
\label{Green}
\ee
For a specific configuration of initial and final points, we can fix: $x=x(j_1)-x(i_1)$,
$y=y(j_1)-y(i_1)$, where the pairs $(x(i_k),y(i_k))$, $(x(j_k),y(j_k))$ label the coordinates of
the vertices $i_k, j_k$ in the sets $I_k$ and $J_k$ respectively. Define $r=\sqrt{x^2+y^2}$. At
$r=0$ the Green function $G_{x,y}$ is singular, and at large distances, $r \gg 1$, the function
$G_{x,y}=G(r)$ has the asymptotic form
\be
G(r)=G(0)-\frac{1}{2\pi}\ln r-\frac{\gamma}{2\pi}-\frac{3\ln 2}{4\pi}+\ldots
\label{GreenLog}
\ee
where $\gamma$ is the Euler constant.

Taking the sets of initial ($I_k$) and final ($J_k$) points as zigzags depicted in the \fig{fig:1}a
and fixing the coordinates of left ends, $i_1 \in I_k$ and $j_1\in J_k$ as $x(i_1)=0, y(i_1)=0$ and
$x(j_1)=x,y(j_1)=y$, we can easily define the coordinates of the vertices $m$ in zigzags $I_k$ and
$J_k$. So, we have
\be
\begin{array}{ll}
\disp x(i_m)=\frac{m}{2}, & \disp \qquad y(i_m)=-\frac{m-2}{2} \medskip \\
\disp x(j_m)=x+\frac{m-2}{2}, & \disp \qquad y(j_m)=y-\frac{m}{2}
\end{array}
\label{vertices1e}
\ee
for $m$ even, and
\be
\begin{array}{ll}
\disp x(i_m)=\frac{m-1}{2}, & \disp \qquad y(i_m)=-\frac{m-1}{2} \medskip \\
\disp x(j_m)=x+\frac{m-1}{2}, & \disp \qquad y(j_m)=y-\frac{m-1}{2}
\end{array}
\label{vertices1o}
\ee
for $m$ odd.

Now we can explicitly write the expression for the determinant $W(I_k,J_k)=\det G(I_k,J_k)$ in
\eq{basic} for specific position of zigzags $I_k$ and $J_k$ ($k$ is odd) shown in the \fig{fig:1}:
\be
W(I_k,J_k)= \det \left(\begin{array}{llllll} G_{x,y} & G_{x,y-1} & G_{x+1,y-1} & G_{x+1,y-2} &
\cdots & G_{x+k,y-k} \bigskip \\  G_{x-1,y} & G_{x-1,y-1} & G_{x,y-1} & G_{x,y-2} & \cdots  &
\bigskip \\
G_{x-1,y+1} & G_{x-1,y} & G_{x,y} & G_{x,y-1} & \cdots &
\bigskip \\
G_{x-2,y+1} & G_{x-2,y} & G_{x-1,y} & G_{x-1,y-1} & \cdots  &
\bigskip \\
\vdots & \vdots & \vdots & \vdots & \ddots &
\bigskip \\
G_{x-k,y+k} &  &  &  &  & G_{x,y}
\end{array}\right)
\label{main_det}
\ee
The generic form of the entry in the position $(m',m)$ of the matrix in \eq{main_det} is
$G_{x_{m,m'},y_{m,m'}}$, where $x_{m,m'}=x(j_m)-x(i_{m'})$, $y_{m,m'}=y(j_m)-y(i_{m'})$,
$\{m,m'\}=1,\ldots, k$.

The explicit expression \eq{main_det} is useful for the determination of the exponent $\nu$ for
small $k=1,3,5,\ldots$. However, the origin of the general expression $\nu=(k^2-1)/2$ remains still
hidden. To reveal it, we note that the bridge construction used in the previous section in the
watermelon definition does not depend on the relative orientation of the sets $I_k$ and $J_k$.
Then, we can chose a more convenient orientation shown in Fig.\ref{fig:1}b. To make the round along
bridges possible, in the latter case we should insert bridges $(i_s,j_{k+1-s})$, $s=1,2,\ldots,k$
instead of former ones $(i_s,j_s)$, $s=1,2,\ldots,k$.

The asymptotic analysis of Eq.(\ref{basic}) does not need exact locations of the points
$i_1,\ldots,i_k$ and $j_1,\ldots,j_k$ in the sites of the square lattice. Instead, we can chose for
$i_1,\ldots,i_k$ and $j_1,\ldots,j_k$ two sets of horizontal coordinates $0\leq u_1<u_2<\ldots,u_k<
{\rm const}$ and $r+v_1<r+v_2<\ldots<r+v_k$ with $0 \leq v_i \leq const$, $i=1,\ldots,k$. The
vertical coordinates of points $i_1,\ldots,i_k$ and $j_1,\ldots,j_k$ will be $0$ for odd indices
and $\xi > 0$ for even ones. Given coordinates $\{u_i\}$ and $\{v_i\}$, the $r$--dependent part of
the determinant in (\ref{basic}) has the asymptotic form
\be
\left|\begin{array}{lllll} \ln[(r-u_1+v_1)^2] & \ln[\xi^2+(r-u_2+v_1)^2]& \ln[(r-u_3+v_1)^2] &
\cdots & \ln[(r-u_k+v_1)^2]
\bigskip \\ \ln [\xi^2+(r-u_1+v_2)^2]& \ln[(r-u_2+v_2)^2] & \ln[(\xi^2+(r-u_2+v_2)^2] & \cdots &
\bigskip  \\ \ln[(r-u_1+v_3)^2] & \ln[\xi^2+(r-u_2+v_3)^2] &  \ln[(r-u_2+v_2)^2]  & \cdots &
\bigskip  \\ \vdots & \vdots & \vdots & \ddots & \bigskip \\
\ln[(r-u_1+v_k)^2] &  &  & & \ln[(r-u_k+v_k)^2]
\end{array}
\right|
\label{matrix}
\ee
The matrix elements in Eq.(\ref{matrix}) are of two types: $\ln[(r-u_i+v_j)^2]$ for odd--odd and
even--even indices $i$ and $j$ and $\ln[\xi^2+(r-u_i+v_j)^2]$ for even--odd and odd--even indices.
For large distance $r$, we have $r\gg u_i$, $r\gg v_i, i=1,\ldots,k$ and $r\gg \xi$. We rewrite the
matrix elements as $2\ln r + 2\ln[1-u_i/r+v_j/r]$ and $2\ln r + \ln[(\xi/r)^2 + (1-u_i/r+v_j/r)^2]$
and expand in powers of $\xi/r$, $u_i/r$ and $v_i/r$.

Consider first the case $\xi=0$. The leading term of the expansion of the determinant \eq{matrix}
is a totally antisymmetric polynomial of the form
\be
\prod_{j>i}^{k}(u_i-u_j)\prod_{j>i}^{k}(v_i-v_j)
\label{lower}
\ee
Each $u_i$ and $v_i$ brings the factor $1/r$, so that the leading term of the expansion is of order
$r^{-k(k-1)}$ ln r.

Consider now the case $\xi>0$. The sets $I_k$ and $J_k$ split into two subsets each, having
vertical coordinates $\xi=0$ and $\xi>0$ correspondingly. To distinguish between different subsets,
we supply coordinates $u_i$ and $v_i$ for even $i$ with hats: $\hat{u_i}$ and $\hat{v_i}$ and
re-enumerate each subset in increasing order. We obtain four sets: $I_{n_1}=\{u_1,u_2,\ldots ,
u_{n_1}\}$, $I_{n_2}=\{\hat{u_1},\hat{u_2},\ldots, \hat{u_{n_2}}\}$,$J_{n_1}=\{v_1,v_2,\ldots,
v_{n_1}\}$, and $J_{n_2}=\{\hat{v_1},\hat{v_2},\ldots,\hat{v_{n_2}}\}$, where $n_1=(k+1)/2$ and
$n_2=(k-1)/2$. The resulting determinant is a totally antisymmetric function with respect to
arguments $u_1,u_2,\ldots,u_{n_1}$, $\hat{u_1},\hat{u_2},\ldots,\hat{u_{n_2}}$, $v_1,v_2,
\ldots,v_{n_1}$ and $\hat{v_1},\hat{v_2},\ldots,\hat{v_{n_2}}$ separately. Therefore, the first
survived term in the expansion of the determinant has the form
\be
\prod_{j>i}^{n_1}(u_i-u_j)\prod_{j>i}^{n_1}(v_i-v_j)\prod_{j>i}^{n_2}(\hat{u_i}-\hat{u_j})
\prod_{j>i}^{n_2}(\hat{v_i}-\hat{v_j})
\label{mix}
\ee
and the contribution from $u,\hat{u},v,\hat{v}$ to the leading term of the expansion is of order
$r^{-\mu}\ln r$, where
\be
\mu=n_1(n_1-1)+n_2(n_2-1)=\frac{k^2}{2}-k+\frac{1}{2}
\label{mu}
\ee
Besides the horizontal coordinates  $u,\hat{u},v,\hat{v}$, we have to take into account the
vertical coordinate $\xi$. The minimal order of $\xi$ in the expansion of the determinant is
$2n_2=k-1$. Indeed, assume that the order of $\xi$ is less than $2n_2$. It means that at least one
row or column of the matrix \eq{matrix} is free of $\xi$, and as consequence, at least one element
of the set $I_{n_2}$ or $J_{n_2}$ should be transferred to $I_{n_1}$ or $J_{n_1}$. Then, we obtain
instead of \eq{mu}
\be
\mu'=\frac{n_1(n_1-1)}{2}+\frac{n_2(n_2-1)}{2}+\frac{n_1(n_1+1)}{2}+\frac{(n_2-2)(n_2-1)}{2}=
\frac{k^2}{2}-k+\frac{5}{2}>\mu
\label{mu'}
\ee
Therefore, the minimal order of $\xi$ is $k-1$ and the resulting exponent in the expansion of the
determinant is
\be
\nu= \mu + k -1 = \frac{k^2-1}{2}
\label{nu}
\ee
The sets $I_k$ and $J_k$ are particular cases of the more general sets $\tilde{I_k}=I_{n_1}\bigcup
I_{n_2}$ and $\tilde{J_k}=J_{n_1}\bigcup J_{n_2}$, so the exponent in \eq{wkrlog} coincides with
$\nu$.

\section{Elongated LERW watermelon}

\subsection{Massive Green functions}

The determinant expression \eq{basic} for the correlation function of a bunch of $k$ loop--erased
random walks remains valid even for stretched spanning trees. When the external elongating field is
applied, the theory becomes massive. That induces an effective lengthscale in the ensemble of
biased loop--erased random walks.

Let $G_{x,y}(\eps)$ be the lattice Green function of a two--dimensional random walk, with the drift
controlled by $\eps$. The function $G_{x,y}(\eps)$ satisfies the following equation (compare to
\eq{laplacian2})
\be
-\left(\frac{1}{4}+\eps\right)\Big(G_{x-1,y}(\eps)+G_{x,y-1}(\eps) \Big) -
\left(\frac{1}{4}-\eps\right)\Big(G_{x+1,y}(\eps)+G_{x,y+1}(\eps) \Big)+
G_{x,y}(\eps)=\delta_{x,0}\delta_{y,0}
\label{diag_drift}
\ee
Define
\be
Q_{x,y}(\eps)=A^{x+y}G_{x,y}(\eps); \qquad A=\sqrt{\frac{1+4\eps}{1-4\eps}}; \qquad \left(0\le \eps
\le \frac{1}{4}\right)
\label{subst}
\ee
The substitution \eq{subst} allows to rewrite \eq{diag_drift} in the symmetric form:
\be
-a\Big\{Q_{x-1,y}(a)+Q_{x,y-1}(a)+Q_{x+1,y}(a)+Q_{x,y+1}(a)\Big\}+Q_{x,y}(a) =
A^{-(x+y)}\delta_{x,0}\delta_{y,0}
\label{symm}
\ee
where $a=\sqrt{1-16 \eps^2}$. By the Fourier transform we get (compare to \eq{Green}):
\be
Q_{x,y}(a)=\frac{1}{8\pi^2} \int_{-\pi}^{\pi}d \alpha \int_{-\pi}^{\pi} d \beta \; \frac{\cos(
\alpha x) \cos(\beta y)}{2-a(\cos \alpha + \cos \beta)}
\label{q}
\ee
Thus, the Green function $G_{x,y}(\eps)$ with the drift $\eps$ is:
\be
G_{x,y}(\eps) = \frac{1}{8\pi^2} \left(\frac{1+4\eps}{1-4\eps}\right)^{\frac{x+y}{2}}
\int_{-\pi}^{\pi} d \alpha \int_{-\pi}^{\pi} d \beta \; \frac{\cos(\alpha x) \cos(
\beta y)}{2-\sqrt{1-16 \eps^2}(\cos \alpha + \cos \beta)}
\label{Green_drift}
\ee
Rewrite $G_{x,y}(\eps)$ as
\be
G_{x,y}(\eps)  = \frac{1}{2} \left(\frac{1+4\eps}{1-4\eps}\right)^{\frac{x+y}{2}} \int_0^{\infty}
d\tau e^{-2\tau} I_x\left(\sqrt{1-16\eps^2}\tau\right) I_y\left(\sqrt{1-16\eps^2}\tau\right)
\label{bessel}
\ee
where $I_x(...)$ and $I_y(...)$ are the modified Bessel functions of integer orders $x$ and $y$
respectively. It is instructive to re-express the equation \eq{bessel} in terms of the
hypergeometric function $_4F_3$:
\be
\begin{array}{l}
\disp G_{x,y}(\eps) = \frac{1}{4} \left(\frac{1}{4}+\eps\right)^{x+y}
\frac{\Gamma(1+x+y)}{\Gamma(1+x)\Gamma(1+y)} \times \medskip \\
\disp \hspace{0.4cm} _4F_3\left(\frac{1+x+y}{2}, \frac{2+x+y}{2}, \frac{1+x+y}{2}, \frac{2+x+y}{2};
1+x+y,1+y,1+x;1-16\eps^2\right)
\end{array}
\label{hyper}
\ee
The equation \eq{hyper}, being an exact expression for the Green function of the random walk on in
open square lattice $(x,y)$ with the drift along the diagonal $x=y$, allows one to analyze the
asymptotic behavior of $G_{x,y}(\eps)$ (at $r=\sqrt{x^2+y^2}\gg 1$) for $\eps\to\frac{1}{4}$
(extreme elongation) and for $\eps\to 0$ (no elongation).

\subsection{Extreme elongation ($\eps\to \frac{1}{4}$)}

To consider the strong elongation, we extract the asymptotic behavior of $G_{x,y}(\eps)$ at
$\eps\to\frac{1}{4}$ and plug this expression into the determinant formula \eq{basic}. The
expression \eq{hyper} allows to find the desired asymptotic behavior of $G_{x,y}(\eps)$ easily.
Introduce the new variables:
\be
z_{\N}=\frac{x+y}{\sqrt{2}}; \quad z_{\T}= \frac{x-y}{\sqrt{2}}
\label{new_xy}
\ee
and rewrite the Green function \eq{hyper} in terms of $z_{\N}$ and $z_{\T}$:
\be
\begin{array}{l}
\disp G_{z_{\N},z_{\T}}(\eps) = \frac{1}{4}\left(\frac{1}{4}+\eps\right)^{\sqrt{2} z_{\N}}
\frac{\Gamma(1+\sqrt{2}
z_{\N})}{\Gamma(1+\frac{z_{\N}+z_{\T}}{\sqrt{2}})\Gamma(1+\frac{z_{\N}-z_{\T}}{\sqrt{2}})} \times
\medskip \\ \disp \hspace{1cm} _4F_3\left(\frac{1}{2}+\frac{z_{\N}}{\sqrt{2}},
1+\frac{z_{\N}}{\sqrt{2}}, \frac{1}{2}+\frac{z_{\N}}{\sqrt{2}}, 1+\frac{z_{\N}}{\sqrt{2}};
1+\sqrt{2} z_{\N},1+\frac{z_{\N}-z_{\T}}{\sqrt{2}},1+\frac{z_{\N}+z_{\T}}{\sqrt{2}};
1-16\eps^2\right)
\end{array}
\label{hyper_new}
\ee
At $\eps\to \frac{1}{4}$ we have
\be
G_{z_{\N},z_{\T}}^{\rm 1D}(\eps) = \frac{2^{-2-\sqrt{2}z_{\N}} \Gamma(1+\sqrt{2}z_{\N})}
{\Gamma(1+\frac{z_{\N}-z_{\T}}{\sqrt{2}})
\Gamma(1+\frac{z_{\N}+z_{\T}}{\sqrt{2}})}\left[1+(1-4\eps)R_1+(1-4\eps)^2R_2 + O(1-4\eps)^3 \right]
\label{series_el}
\ee
where $R_1\equiv R_1(z_{\N},z_{\T})$ and $R_2\equiv R_2(z_{\N},z_{\T})$ are rational functions of
$z_{\N}$ and $z_{\T}$:
\be
\begin{cases}
R_1(z_{\N},z_{\T})=\frac{2+z_{\N}^2+\sqrt{2}z_{\N}(2+z_{\T}^2)}
{2(2+2\sqrt{2}z_{\N}+z_{\N}^2-z_{\T}^2)} \medskip \\
R_2(z_{\N},z_{\T})=\frac{6 z_{\N}^4+\sqrt{2} z_{\N}^3 \left(4 z_{\T}^2+35\right)+2 z_{\N}^2 \left(2
z_{\T}^4+18 z_{\T}^2+73\right)-2 \sqrt{2} z_{\N} \left(z_{\T}^4-22 z_{\T}^2-64\right)+8
\left(z_{\T}^2+10\right)}{16 \left(z_{\N}^4+6 \sqrt{2} z_{\N}^3-2 z_{\N}^2
\left(z_{\T}^2-13\right)-6 \sqrt{2} z_{\N} \left(z_{\T}^2-4\right)+z_{\T}^4-10 z_{\T}^2+16\right)}
\end{cases}
\label{rational}
\ee

Denoting $\omega=\frac{1}{4}-\eps$ and applying the Stirling formula to $\ln G_{z_{\N},z_{\T}}^{\rm
1D}$ at $z_{\N}\gg 1$, we get the following asymptotic behavior at large separations:
\be
\ln G_{z_{\N},z_{\T}}^{\rm 1D}=\ln \frac{1+2 \omega + 6\omega^2}{2^{7/4}\sqrt{\pi}} -\frac{1}{2}\ln
z_{\N} -\frac{1+4z_{\T}^2+2\omega(1-4 z_{\T}^2)+
2\omega^2(7-4z_{\T}^2)}{4z_{\N}\sqrt{2}(1+2\omega+6\omega^2)} +
O\left(\frac{1}{z_{\N}}\right)^{3/2}
\label{gauss}
\ee
Keeping only the first order terms in $\omega=\frac{1}{4}-\eps$, we get from \eq{gauss} the
following asymptotic behavior for the Green function $G_{z_{\N},z_{\T}}^{\rm 1D}$ at large
separations:
\be
G_{z_{\N},z_{\T}}^{\rm 1D} \approx \frac{(1+2\omega)}{2^{7/4}\sqrt{\pi
z_{\N}}}e^{-\frac{(1-2\omega)z_{\T}^2}{\sqrt{2}(1+2\omega)z_{\N}}}\approx
\frac{1}{2^{7/4}\sqrt{(1-4\omega) \pi z_{\N}}}e^{-\frac{(1-4\omega)z_{\T}^2}{\sqrt{2}z_{\N}}}
\label{gauss2}
\ee
Thus, we see that \eq{gauss2} has a form of a Gaussian distribution for a one--dimensional random
walk of length $z_{\N}$ and the transversal coordinate $z_{\T}$ for any strong elongations
$0<\omega = \frac{1}{4}-\eps \ll 1$. The separation distance $z_{\N}$ between two ladders shown in
the \fig{fig:1} can be interpreted as a length of a directed trajectory.

For $\varepsilon=\frac{1}{4}$, the Green function $G_{z_{\N},z_{\T}}^{\rm 1D}$ becomes the
probability to visit point $x=z_{\T}$ of the one-dimensional lattice $\{x \in Z^1\}$ by the
symmetric random walk after $z_{\N}$ time steps, provided that it is located at the origin $x=0$ at
the initial moment of time $z_{\N}=0$. The sets $I_k$ and $J_k$ in the determinant formula
\eq{basic} can be interpreted in terms of variables $z_{\N}$ and $z_{\T}$ simply as groups of $k$
neighboring sites of the one-dimensional lattice taken at the moments of time $t=0$ and $t=z_{\N}$.
Then, formula \eq{basic} takes the form of the Karlin--McGregor formula (known also as the
Lindstrom--Gessel--Viennot formula and used by Huse and Fisher for calculation of the reunion
probability of the vicious walkers). The result obtained in \cite{Fisher, HuseFisher} for large
number of steps $t$ is given by \eq{wkrFisher} where $r=z_{\N}=t$.

\subsection{Arbitrary elongation $\eps>0$}

The direct expansion of \eq{hyper} for small $\eps$ and large $x,y$ meets essential technical
difficulties. Taking into account that for $x\gg 1$ and $y\gg 1$ the main contribution to the
integral in \eq{Green_drift} is given by values $|\alpha|<\frac{1}{x}$ and $|\beta|<\frac{1}{y}$.
Thus, at large separations $r\gg 1$ and at small $\eps$, we can use the saddle--point
approximation. To proceed coherently let us replace the discrete recursion relation \eq{diag_drift}
by the continuous one
\be
-2\eps\left(\partial_x+\partial_y\right)G_{x,y}(\eps) +
\frac{1}{4}\left(\partial^2_{xx}+\partial^2_{yy} \right)G_{x,y}(\eps) = \delta_{x,y}
\label{diag_cont}
\ee
Note that the equation \eq{diag_cont} is valid for any $r$ and $\eps$. The solution to
\eq{diag_cont} is
\be
G_{x,y}(\eps) = -\frac{e^{4\eps(x+y)}}{4\pi^2} \int_{-\infty}^{\infty} d \alpha
\int_{-\infty}^{\infty} d \beta \; \frac{e^{i (x \alpha + y \beta)}} {8\eps^2 +
\frac{1}{4}(\alpha^2 + \beta^2)}
\label{asympt}
\ee
The expression \eq{asympt} can be easily evaluated exactly. Introducing the polar coordinates
$$
\alpha = \rho \cos \phi; \quad  \beta = \rho \sin \phi
$$
rewrite \eq{asympt} as
\be
G_{x,y}(\eps) = -\frac{e^{4\eps(x+y)}}{4\pi^2} \int_0^{\infty} \rho d \rho \int_0^{2\pi} d \phi \;
\frac{e^{i \rho (x \cos \phi + y \sin \phi)}} {8\eps^2 + \frac{1}{4}\rho^2}
\label{polar}
\ee
Since
$$
x \cos \phi + y \sin \phi = \sqrt{x^2+y^2} \cos(\phi-\gamma)
$$
where $\gamma = \arccos \frac{x}{\sqrt{x^2+y^2}}$, we get:
\be
G_{x,y}(\eps) = -\frac{e^{4\eps(x+y)}}{\pi} \int_0^{\infty} \rho d \rho \frac{J_0\left(\rho
\sqrt{x^2+y^2}\right)}{16\eps^2 + \frac{1}{2}\rho^2} = -\frac{2e^{4\eps(x+y)}}{\pi}
K_0\left(4\sqrt{2}\eps\sqrt{x^2+y^2}\right)
\label{polar2}
\ee
Rotating the coordinates by $\pi/4$, i.e. passing to $(z_{\N},z_{\T})$ defined in \eq{new_xy}, we
arrive at the final expression
\be
G_{x,y}(\eps) = -\frac{2}{\pi}e^{4\sqrt{2}\eps z_{\N}} K_0\left(4\sqrt{2}\eps
\sqrt{z_{\N}^2+z_{\T}^2}\right)
\label{polar3}
\ee
To extract the asymptotic behavior of $G_{x,y}(\eps)$ at any $\eps>0$ and large $z_{\N}$, we should
first expand logarithm of \eq{polar3} ar $z_{\N}\gg 1$ and then consider the elongations with any
$\eps>0$:
\be
\ln G_{x,y}(\eps) = -\ln \left(2^{3/4} \sqrt{\pi \eps} \right) - \frac{1}{2} \ln z_{\N} -
\frac{1+128 \eps^2 z_{\T}^2}{32 \sqrt{2} \eps z_{\N}} + O\left(\frac{1}{z_{\N}^{3/2}} \right)
\label{log_big}
\ee
Thus, we get from \eq{log_big} at $z_{\N}\gg z_{\T}\gg \eps^{-1}$:
\be
G_{x,y}(\eps)\approx \frac{1}{2^{3/4} \sqrt{\pi \eps z_{\N}}}e^{-\frac{4 \eps
z_{\T}^2}{\sqrt{2}z_{\N}}}
\label{big}
\ee

In Appendix we demonstrate that similar expression can be obtained for stretching along the
$x$--axis. To demonstrate the generality of the result, we perform the computations for the
discrete model.

\section{Elongated LERW watermelon vs vicious walkers}

The explicit expression for the determinant (\ref{basic}) is
\be
W(I_k,J_k)=\sum_{\pi \in S_k}\delta(\pi ) G_{1,\pi(1)}G_{2,\pi(2)}\ldots G_{k,\pi(k)}
\label{expandet}
\ee
where $\delta(\pi)$ is the sign of permutation $\pi$ and $G_{m,n}$ is the Green function
$G_{x_{m,n},y_{m,n}}$ defined in (\ref{main_det}).

The expressions (\ref{gauss2}) and (\ref{big}) allow us to write (\ref{expandet}) in the form
\be
W(I_k,J_k)=\sum_{\pi\in S_k}\delta(\pi)\, \frac{e^{-\frac{|{\bf z}-\pi{\bf z}^0|^2}{2Dr}}} {(2\pi D
r)^{k/2}}
\label{Gaussdet}
\ee
The diffusion constant $D\equiv D(\eps)$ depends on elongation $\eps$ and defines the mean--square
displacement $\la(z_i-z^0_i)^2\ra = Dr$ in the transverse direction for the ``time'' $r$. The
coordinates $z_1,\ldots z_k$ and $z_1^0,\ldots z_k^0$ of vectors ${\bf z}$ and ${\bf z}^0$ denote
the positions of points $i_1,\ldots i_k$ and $j_1,\ldots j_k$ measured along corresponding zigzags.
The expression (\ref{Gaussdet}) coincides with that used by Huse and Fisher (see Appendix of
\cite{HuseFisher}) for the derivation of asymptotic reunion probability of $k$ vicious walkers.

The computation of \eq{Gaussdet} is based on the analysis of antisymmetric properties of variables
$z_1,\ldots z_k$ and $z_1^0,\ldots z_k^0$ as well as on the computation of \eq{matrix} was based on
the analysis of antisymmetric properties of variables $u_1,\ldots u_k$ and $v_1,\ldots v_k$. It is
instructive to compare procedures leading to isotropic and elongated asymptotic behavior.

Using the invariance  $|\pi {\bf z}^0|=|{\bf z}|$, one can write (\ref{Gaussdet}) as
\be
W(I_k,J_k)= U_k({\bf z},{\bf z}^0)\frac{e^{-\frac{|{\bf z}|^2+|{\bf z}^0|^2}{2Dr}}}{(2\pi D
r)^{k/2}}
\label{modGaussdet}
\ee
where
\be
U_k({\bf z},{\bf z}^0)=\sum_{\pi\in S_k}\delta(\pi)e^{\frac{{\bf z}\,\pi{\bf z}^0}{2Dr}}
\label{U}
\ee
Since the length of zigzags $L$ is much smaller than the separation between clusters $I$ and $J$,
one can expand the exponential in powers of $\frac{{\bf z}\,\pi{\bf z}^0}{2Dr}$. Each term of the
expansion is a homogeneous polynomial in $k$ variables $z_i$ or, equivalently, $z^0_i$. The lowest
order of polynomials is of degree $k(k-1)/2$ due to the requirement of total antisymmetry. Each
$z_i$ or $z^0_i$ brings the factor $1/r$, so the asymptotic behavior of $U_k$ is $U_k \sim 1/r
^{k^2/2-k/2}$. Combining this factor with $1/(2\pi D r)^{k/2}$ in \eq{modGaussdet}, one gets the
asymptotic reunion probability \eq{wkrFisher}, $W_k(r) \sim r^{-\frac{k^2}{2}}$.

The first difference between the isotropic and elongated watermelons is in the dependence of the
determinant expansion in $u_i$ and $v_i$ for isotropic watermelons and in $z_i$ and $z^0_i$ for
stretched ones. The arguments $z_i$ and $z^0_i$ enter into the expansion pairwise, while the
expansion of \eq{matrix} depends on $u_i$ and $v_i$ independently, so both $u_i$ and $v_i$ bring
the factor $1/r$ each.

The second difference is that matrix elements of \eq{expandet} are uniform, while those of
\eq{matrix} contain the additional argument $\xi$ depending on the parity of indices. Each $\xi$
increases the power of $1/r$ in the expansion by one (see the explanations at the end of the
Section III).

Finally, the determinant  \eq{Gaussdet} contains a ``diffusive'' factor $1/(2\pi D r)^{k/2}$
originating from the asymptotic form of the Green function \eq{gauss2}. Being collected all
together, these factors lead to different laws \eq{eq:6} and \eq{wkrFisher} for isotropic and
elongated watermelons.

\section{Discussion}

\subsection{Reunion exponents for LERW and VW watermelons}

Using the combinatorial approach based on exact enumeration of spanning trees on a square lattice,
we have reproduced the Karlin--McGregor (or Lindstrom--Gessel--Viennot formula for vicious random
walks, or Slater determinant for free fermions) from the determinant expression for reunion
probability of loop--erased stretched watermelons.

Let us recall the Karlin--McGregor formula \cite{karlin}. Consider $k$ (1+1)--dimensional Markov
chains $\zeta_1(t),..., \zeta_k(t)$ with $\zeta_i(t=0) = x_i^{(1)}$ and $\zeta_i(t=T) = x_i^{(2)}$,
for all $i=1,...,k$. Assume that the time, $t$, is homogenous and any two walks cannot touch each
other. Then the probability $P_T\{\zeta_1(t)<\zeta_2(t)<...<\zeta_k(t)\}$, $(0\le t \le T)$ to have
a bunch of $k$ uncrossing random walks each of length $T$ with starting points $\{\zeta(0)\}$ and
ending points $\{\zeta(T)\}$, is given by the following expression (Karlin--McGregor formula)
\be
P_T\{\zeta_1(t)<\zeta_2(t)<...<\zeta_k(t)\} = \det\left[P_T\left(\zeta_i(0),\zeta_j(T)\right)_{1
\le i,j \le n}\right]
\label{karlin}
\ee
where $P_T\left(\zeta_i(0),\zeta_j(T)\right)$ is the probability for a single (1+1)--dimensional
random walk of length $T$ to have two extremities at the points $\zeta_i(0)$ and $\zeta_j(T)$
($1\le i,j \le k$).

In the original determinant expression \eq{basic} (for $\eps=0$) we have no any time scale in the
system since the building block $G_{x,y}$ of the matrix $W$ in \eq{basic} is the logarithmically
decaying propagator of the 2D free field. However, when the bias $\eps>0$ is switched on, the
theory becomes massive, what signals the occurrence of a finite length scale in the system,
$\ell\sim \eps^{-1}$. The corresponding partition function, $G_{x,y}$, becomes Gaussian at strong
elongations controlled by the parameter $\eps$ (see Eq.\eq{gauss2}). Thus, the equation
\eq{main_det} initially written for a $k$--leg symmetric watermelon of two--dimensional loop erased
random walks, converts at ($0<\frac{1}{4}-\eps \ll 1$) into the Karlin--McGregor formula for a
bunch of $k$ (1+1)--dimensional vicious random walks.

The value of the reunion exponents $\nu=\frac{k^2-1}{2}$ for the 2D $k$--leg watermelon transforms
at strong elongations into the (1+1)D value $\tilde{\nu}=\frac{k^2}{2}$. Note, that $\tilde{\nu}$
differs by the constant $-\frac{1}{2}$ from a reunion exponent $\nu'=\frac{k^2-1}{2}$ of the (1+1)D
vicious random walks. The difference between $\tilde{\nu}$ and $\nu'$ deals with the fact that the
value $\tilde{\nu}$ obtained from Karlin--McGregor formula imply that the end points of the bunch
meet in some {\em specific} point, while to get $\nu'$ we should allow for a reunion in {\em every}
point in the space, i.e. intergrate over the position of the end point of the bunch.

\subsection{Geometry of elongated watermelons and dimers}

The elongation of spanning trees and the passage from isotropic to directed watermelons made of
loop--erased random walks, has a transparent geometric interpretation and can be understood in
terms of configurations of dimers on square and honeycomb lattices.

\begin{figure}[ht]
\epsfig{file=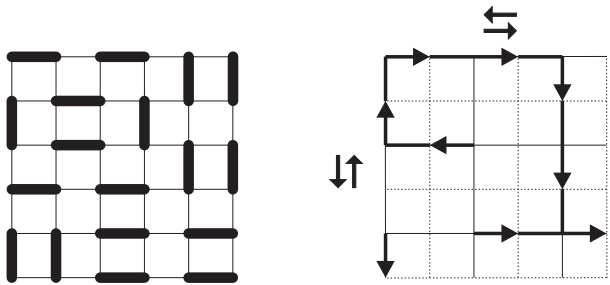,width=8cm}
\caption{Configuration of dimers on the square lattice and corresponding configuration of symmetric
spanning tree on the odd--odd square sublattice.}
\label{fig:03}
\end{figure}

It is known (see, for example, \cite{priezzhev-dim}) that there is a bijection between the
configuration of dimers on a square lattice $\mathbb{Z}_2$ and the a spanning forrest on, say, an
odd--odd sublattice $\mathbb{Z}_2^{+}$ of $\mathbb{Z}_2$. This bijection is set by the construction
described in \cite{priezzhev-dim}. In brief, this construction is as follows. Consider the dimers
located on the odd--odd sublattice $\mathbb{Z}_2^{+}$ only. Replace the dimers by vectors whose
starting points are in the vertices of this sublattice. Continue these vectors until meeting the
next neighboring dimer (vector) -- as it is schematically shown in the \fig{fig:03}. Erasing
arrows, we get on $\mathbb{Z}_2^{+}$ the spanning forrest, which is in bijection with the initial
configuration of dimers. The configuration of dimers on the even--even sublattice
$\mathbb{Z}_2^{-}$ can be uniquely restored since dimers fully cover the square lattice
$\mathbb{Z}_2$ without common points and empty vertices.

\begin{figure}[ht]
\epsfig{file=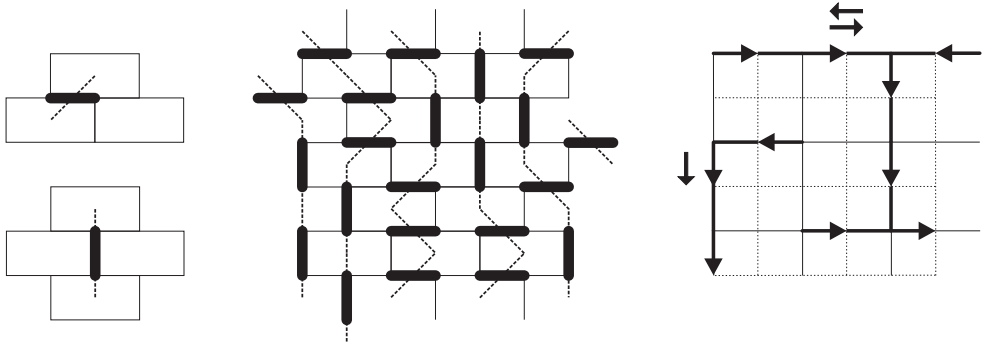,width=13cm}
\caption{Illustration of bijection between dimers on the honeycomb lattice and fully elongated
spanning forrest along $y$ axis on an odd--odd square sublattice.}
\label{fig:04}
\end{figure}

The same algorithm we can apply to dimers on a honeycomb lattice (drawn in a form of a brick
lattice). This establishes the bijection between dimers on a brick lattice and {\em elongated}
spanning forest on an odd--odd square sublattice $\mathbb{Z}^{+}_2$ -- see \fig{fig:04}. Thus, the
fully elongated spanning forrest (along the vertical axis in \fig{fig:04}) is in one--to--one
correspondence with configurations of dimers on the honeycomb lattice. According to the
correspondence, discovered by \cite{nagle} (we have reproduced it schematically in the
\fig{fig:04}), the configurations of dimers on a honeycomb (brick) lattice is in bijection with
ensemble of directed mutually avoiding lattice random walks (``vicious'' walks).

As we have seen in the previous Section, the extreme elongation of loop--erased watermelons on the
square lattice brings us to statistics of vicious walks extremely stretched along the diagonal of
the square lattice. The geometric connection between spanning forest on a square and on a honeycomb
lattices is consistent with this results (up to the rotation of the direction of extension by
$\pi/4$).

\subsection{From CFT to RMT}

We would like to pay attention to the following observation. We see that the critical exponent
$\nu_{\rm 2D}$ for $\eps = 0$ defines the scaling behavior of the $k$--leg correlation function of
simplectic fermions described by $c=-2$ logarithmic conformal field theory and the reunion
probability is given by the determinant expression \eq{main_det}. For $0<\eps\le \frac{1}{4}$ the
theory is not conformal anymore because of the presence of the mass. At strong drifts, $\eps
\approx \frac{1}{4}$ the system becomes (1+1)--dimensional with the critical exponent
$\tilde{\nu}=\frac{k^2}{2}$ and the reunion probability is described by the Karlin--McGregor
formula which can be straightforwardly derived from \eq{main_det} by taking $\eps \to\frac{1}{4}$
limit.

The interpolation between the determinant \eq{main_det} and the Karlin--McGregory \eq{karlin}
formulas as $\eps$ is changing from $0$ to $\frac{1}{4}$, allows us suggest the possibility to pass
in the framework of unified description from the logarithmic conformal field theory to the random
matrix theory since at extreme elongations the bunch of directed watermelons (vicious random walks)
represents the bunch of interacting Dyson random walks \cite{dyson} described by the Gaussian
Random Matrix Ensemble.

\subsection{From LERW to VW via integrability}

According to \cite{cardy}, the isotropic LERW in the continuum limit is described by $SLE_k$
corresponding to the conformal field theory with the central charge $ c=1-\frac{24}{k}
\left(\frac{k^2}{4}-1\right)^2$. For $k=2$ one arrives at logarithmic CFT with $c=-2$. Let us use
now the relation between $SLE_2$ and the trigonometric Calogero system found by Cardy in
\cite{cardy}. He has shown that the stochastic radial SLE evolution is governed by the
trigonometric Calogero Hamiltonian
\be
H_{cal}=-\frac{1}{2}\sum_{j=1}^N \frac{\partial^2}{\partial x_j^2} + \frac{\beta(\beta-2)}{16}
\sum_{j<k} \sin^{-2}\frac{x_j-x_k}{2}
\ee
where $\beta=\frac{8}{k}$. For $k=2$ the coupling constant corresponds to the symplectic ensemble,
$\beta=4$. The $n$--particle Calogero wave function depends on the position of the SLE roots on the
circle and corresponds to the probability in the LERW model. Simultaneously, the wave function can
be identified with the particular conformal block in the $c=-2$ model
\be
\Psi_{cal}(z_1,\dots z_n)\propto <\Phi(0)V_{2,1}(z_1)\dots V_{2,1}(z_n)>
\ee
where $V_{2,1}$ is associated with the states degenerate at the second level.

The observation by Cardy could be used for the interpolation between different stochastic systems.
Indeed, there exists an expected correspondence between the stochastic processes and various
quantum Hamiltonian systems, so the Calogero model is not an exception. Since no elliptic
parameters (or, equivalently, nonperturbative effects) involved, the most general relevant
integrable many--body system is the trigonometric Ruijsenaars model which is the
relativistic/discrete generalization of the Calogero system. The wave function of the trigonometric
Ruijsenaars model is identified with the Macdonald polynomial. The corresponding stochastic
counterpart, the Macdonald process, parameterized by $(q,t)$, has been introduced recently in
\cite{borodin}.

\begin{figure}[ht]
\epsfig{file=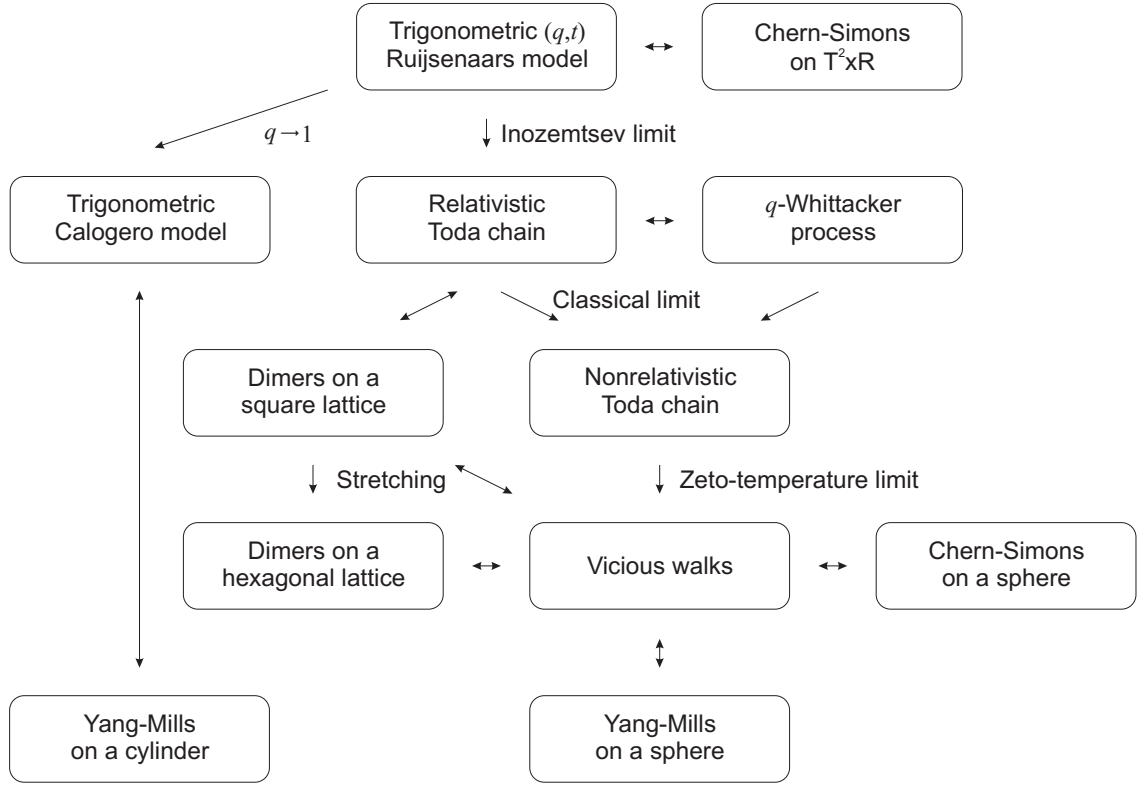,width=15cm}
\caption{The flowchart of possible degenerations of the trigonometric Ruijsenaars model.}
\label{fig:list}
\end{figure}

There are several degenerations of the Macdonald process \cite{borodin}. The simplest one $q\to 1$
brings us back to the trigonometric Calogero model. A more involved, so-called Inozemtsev scaling
limit \cite{inozemtsev}, provides the relativistic Toda system or, the $q$--Whittacker process, in
the terminology of \cite{borodin}. Further degeneration yields the usual Toda system or Whittacker
process introduced by O'Connell \cite{ocon1,ocon2}, which has been found in several models other
(see also \cite{katori, nechaev}).
Let us emphasize that the coupling constant $t$ in the rational Ruijsenaars model fixes in general
situation the central charge of the Virasoro algebra similar to the Cardy's case. The list of
mentioned degenerations is schematically depicted in the figure \fig{fig:list}.

Turning back to the anisotropic lattice LERW model considered in the paper we could ask about its
integrable counterpart. Since the isotropic continuum LERW correspond to the trigonometric Calogero
model at fixed coupling constant, we should find its generalization in two direction because we
have now anisotropy and the lattice size parameters.

\noindent {\bf Conjecture}. Since the lattice parameter implies the relativization of the Calogero
model, the anisotropic LERW is mapped onto the relativistic Calogero model with an arbitrary value
of coupling constant (i.e. of central charge).

What could be the evidences in favor of such identification?

\begin{itemize}

\item On the one hand, looking from the LERW side, we know that LERW model is closely related to
the dimer model on the squire lattice. The integrable system associated to dimers, is the
relativistic Toda chain \cite{franco}, which itself is the limit of the trigonometric Ruijsenaars
model, as we expect. However yet it is not clear how the Inozemtsev limit is set up in terms of
dimers.

\item On the other hand, we expect that the integrable counterpart of discrete vicious walks is the
model of free relativistic fermions involved into the Chern--Simons action. Indeed it has been
found in \cite{deHaro, forrester} that the reunion probability in VW model is identified with the
2D Yang--Mills theory on the sphere. It is well--known that lifting YM theory to the perturbed
Chern--Simoms with the same group gives the relativization of the system, where the level of the
Kac--Moody algebra provides the relativization/discretization parameter. The CS theory degenerates
to 2D Yang--Mills in the $k\to \infty$ limit.

\end{itemize}

It is useful to have in mind that the trigonometric Ruijsenaars model emerges from the perturbed
Chern--Simons model on the torus with the marked point \cite{gn2}, while the trigonometric Calogero
model corresponds to perturbed 2D Yang--Mills theory on the cylinder \cite{gn1}. Hence the
stochastic process is formulated in terms of the particular observables in the perturbed
topological gauge theories. The coordinates of the particles involved are obtained from the
holonomies of the gauge connections over the cycles upon the $T$--duality transformation. In
particular, coordinates of the wave function of the Calogero model, which can be visualized via the
cross--section of the watermelon in the middle, corresponds to the holonomies in the 2D Yang--Mills
model. Note that we start with the discrete model in 2D hence at the extreme elongation we get the
(1+1) system with the discrete time dynamics.

The relation with the integrable models suggests another analogy which could be of some use. It is
known that the Calogero model is related to the Quantum Hall Effect(QHE) at the continuum. The
external magnetic field defines the coupling constant in the Calogero model. The type of Calogero
model depends on the geometry of the QHE system.  Suppose that we are in the limit of the strong
magnetic field. Then the degrees of freedom are localized at the lowest Landau level and the 2D
system becomes effectively (1+1) dimensional where the time is identified with the radial
coordinate on the plane. This reduction is very similar to the one considered in the paper, however
the decomposition of 2D into (1+1) is different.

It would be interesting to see if two different conformal behaviors similar to discussed above
could be found in QHE. Since the interpolation parameter is the magnetic field strength, we could
ask what happened when we change it. The Calogero system describes the gapless boundary modes of
the QHE droplet, corresponding to the CFT with the central charge fixed by the filling fraction
(or, equivalently by the magnetic field). What is the c=-2 counterpart in the case of ``magnetic
elongation''? Fortunately the answer is known: and the logariphmic conformal field theory describes
the transition between the two QHE platoes. Hence, when the strength of the magnetif field is
increasing, one evolves from one minimal CFT to another with the LCFT critical behavior. It seems
very interesting to pursue further this analogy further.

\begin{acknowledgments}

We are grateful to R. Santachiara and G. Schehr for valuable discussions on all stages of the work
and to S.N. Majumdar and H. Saleur for useful suggestions. The work of A.G. is supported in part by
grants RFBR-12-02-00284 and PICS-12-02-91052; the work of S.N. is supported by the ANR grant
2011-BS04-013-01 WALKMAT; the work of V.B.P. is supported by grants RFBR 12-01-00242, 12-02-91333
and the DFG grant RI 317/16-1

\end{acknowledgments}

\begin{appendix}
\section{Extension along an axis: Saddle point for discrete model}

Consider the Green function $G$ of the Laplacian $\Delta$ on the anisotropic square lattice defined
as a solution of the equation
\be
4 G_{x,y} - p_1 G_{x-1,y} - q_1  G_{x+1,y} - p_2 G_{x,y-1} - q_2  G_{x,y+1} = \delta_{x,0}
\delta_{y,0}
\label{general_drift}
\ee
Here the parameters of anisotropy (drift) $p_1$, $q_1$, $p_2$ and $q_2$ are positive and satisfy
the identity
\be
p_1 + q_1 + p_2 + q_2 = 4
\ee
It is easy to see that the integral representation of the Green function is
\be
G_{x,y} = \frac{1}{4\pi^2}\int_{-\pi}^{\pi} d\alpha \int_{-\pi}^{\pi} d\beta \frac{ e^{ - i x
\alpha} e^{ - i y \beta} } {4 - p_1 e^{ i \alpha} - q_1 e^{ -i \alpha} - p_2 e^{ i \beta} - q_2 e^{
-i
\beta}}
\ee
Let us introduce the notations
\be
\xi_1 = \sqrt{p_1 q_1}\,, \quad \xi_2 = \sqrt{p_2 q_2}\,, \quad \gamma_1 =
\sqrt{\frac{q_1}{p_1}}\,, \quad \gamma_2 = \sqrt{\frac{q_2}{p_2}}
\ee
Changing the variables $\alpha$ and $\beta$
\be
e^{i\alpha} \to \gamma_1 e^{ i\alpha}, \quad e^{i\beta} \to \gamma_2 e^{ i \beta}
\ee
we get
\be
G_{x,y} = \frac{1}{8\pi^2 \gamma_1^x \gamma_2^y}  \int_{\pi}^{\pi} d \alpha \int_{-\pi}^{\pi} d
\beta\; \frac{e^{- i x \alpha} e^{ -i y \beta} } {2 - \xi_1\cos\alpha - \xi_2\cos\beta}
\ee
Integrating over $\alpha$, we arrive at the following expression:
\be
G_{x,y} = \frac{1}{2\pi \left(\xi_1\gamma_1\right)^x \gamma_2^y} \int_{0}^{\pi} d\beta \;
\frac{\left(2 - \xi_2 \cos\beta-\sqrt{(2 - \xi_2 \cos\beta)^2 - \xi_1^2}\right)^{x} \;
\cos(y\beta)}{\sqrt{(2 - \xi_2\cos\beta)^2 - \xi_1^2}}
\label{saddle}
\ee
The asymptotics of the Green function for fixed finite $y$ and $x \gg |y|$ can be obtained using
the saddle point approximation. To proceed, expand the exponent in \eq{saddle} near the maximum
$\beta=0$.
\begin{multline}
\left( 2 - \xi_2 \cos\beta - \sqrt{( 2 - \xi_2 \cos\beta )^2-\xi_1^2} \right)^x =
\left( 2 - \xi_2 -\sqrt{(2 - \xi_2)^2-\xi_1^2} \right)^x \medskip \\
\times \exp\left\{-\frac{\xi_2\, x\, \beta ^2 }{2 \sqrt{(2-\xi_2)^2 - \xi_1^2}}\right\} \left( 1 +
O(\beta^4) \right)
\end{multline}
The denominator of the integrand can be expanded in series in $\beta$ and integrated term by term.
For the main contribution, we have
\be
G_{x,y} \simeq \frac{\left(2 - \xi_2 - \sqrt{(2 - \xi_2)^2 - \xi_1^2}\right)^x} {2\pi
\left(\xi_1\gamma_1\right)^x \gamma_2^y \sqrt{(2 - \xi_2)^2 - \xi_1^2}} \int_{0}^{\infty} d \beta\;
e^{-\frac{\xi_2\, x\, \beta ^2}{2 \sqrt{(2-\xi_2)^2 - \xi_1^2}}} \cos(y\beta)
\label{series}
\ee
which gives
\be
G_{x,y} \simeq \frac{e^{-\frac{y^2\sqrt{(\xi_2-2)^2-\xi_1^2}}{2 x \xi_2}}}{ 2 \sqrt{2 \pi \xi _2}
\gamma_2^y \left[\left(2-\xi_2\right)^2 - \xi_1^2\right]^{1/4}\sqrt{x}} \left(\frac{2 - \xi_2 -
\sqrt{\left(2-\xi_2\right)^2 - \xi_1^2} }{ \xi_1 \gamma_1 }\right)^x, \quad x \gg |y|
\label{series1}
\ee
Next terms of $G_{x,y}$ can be found by keeping higher powers in $\beta$ and integrating
sequentially.

Assuming in \eq{series} that $p_2=q_2=1$, i.e. $\xi_2=\gamma_2=1$, and $p_1=1 + \delta$, $q_1=1 -
\delta$ with $\delta>0$, we get
\be
G_{x,y} \simeq \frac{1}{2\sqrt{2\pi}\sqrt{\delta} \sqrt{x}}\; e^{-\frac{y^2 \delta}{2 x}} \;,\, x
\gg |y|.
\label{el_asymp}
\ee
Next terms of the asymptotic expansion of the Green function for the symmetric walk in vertical
direction ($p_2=q_2=1$) and fully elongatated in horizontal direction ($\delta=1$) are as follows
\be
G_{x,y} \simeq \frac{1}{2 \sqrt{2 \pi} \sqrt{x}} \left( 1 -\frac{y^2}{2 x} + \frac{y^4-4 y^2+1}{8
x^2} - \frac{4 y^6 - 80 y^4 + 196 y^2 + 1495}{192 x^3} + \cdots \right)
\ee

\end{appendix}


\begin{thebibliography}{99}

\bibitem{Kirhhoff} G. Kirchhoff, Ann. Phys. Chem. {\bf 72}, 497 (1847).

\bibitem{SalDup} H.Saleur and B. Duplantier, Phys. Rev. Lett. {\bf 58}, 2325 (1987).

\bibitem{Pri94} V.B. Priezzhev, J. Stat. Phys. {\bf 74},955 (1994).

\bibitem{Bak} P.Bak, C. Tang, and K. Wiesenfeld, Phys. Rev. Lett. {\bf 59}, 381 (1987).

\bibitem{Dhar} D.Dhar, Phys. Rev. Lett. {\bf 64}, 1613 (1990).

\bibitem{Fisher} M.E.Fisher, J. Stat. Phys. {\bf 34}, 667 (1984).

\bibitem{HuseFisher} D.A.Huse and M.E. Fisher, Phys. Rev. B {\bf 29}, 239 (1984).

\bibitem{Lawler} G.F. Lawler, O. Shram, and W. Werner, Ann. Probab. {\bf 32}, 939 (2004)

\bibitem{Wilson} J. G. Propp and D.B. Wilson, J. of Algorithms, {\bf 27}, 170 (1998).

\bibitem{Iva} E.V. Ivashkevich and Chin-Kun Hu, Phys. Rev. E {\bf 71}, 015104 (R) (2005).

\bibitem{KtitIvaPri} E.V. Ivashkevich, D.V. Ktitarev, and V.B. Priezzhev, J. Phys. A: Math. Gen.
{\bf 27}, L585 (1994).

\bibitem{MajDhar} S.N. Majumdar and D. Dhar, J. Phys. A: Math. Gen. {\bf 24}, L357 (1991).

\bibitem{karlin} S. Karlin and J. McGregor, Pacific J. Math. {\bf 9}, 1141 (1959).

\bibitem{priezzhev-dim} Priezzhev, Soviet Physics Uspekhi {\bf 28}, 1125 (1985).

\bibitem{nagle} J.F. Nagle, Phys. Rev. Lett. {\bf 34}, 1150 (1975).

\bibitem{dyson} F. J. Dyson, J. Math. Phys. {\bf 3}, 1191 (1962).

\bibitem{cardy} J.L. Cardy,
  Phys. Lett. B {\bf 582}, 121 (2004) [hep-th/0310291].

\bibitem{deHaro} S. de Haro and M. Tierz,
  Phys. Lett. B {\bf 601}, 201 (2004) [hep-th/0406093];
 S. de Haro,
  JHEP {\bf 0408}, 041 (2004) [hep-th/0407139].

\bibitem{forrester}  P.J. Forrester, S.N. Majumdar, and G. Schehr,
  Nucl. Phys. B {\bf 844}, 500 (2011) [Erratum-ibid. B {\bf 857}, 424 (2012)]
  [arXiv:1009.2362 [math-ph]].

\bibitem{borodin}  A. Borodin and I. Corwin, arXiv:1111.4408

\bibitem{inozemtsev} Inozemtsev, Comm. Math. Phys., {\bf 121} 629 (1989).

\bibitem{ocon1} N. O'Connell, Ann. Probab. {\bf 40}, 437 (2012) [arXiv:0910.0069].

\bibitem{ocon2} N. O'Connell, arXiv:1201.4849

\bibitem{katori} M. Katori, Phys. Rev. E {\bf 84}, 061144/1-11 (2011) [arXiv:1110.1845]

\bibitem{katori} M. Katori, J. Stat. Phys. {\bf 147}, 206 (2012) [arXiv:1112.4009].

\bibitem{nechaev} A. Gorsky, S. Nechaev, R. Santachiara, and G. Schehr, Nuclear Physics B
{\bf 862} [FS], 167 (2012) [arXiv:1110.3524].

\bibitem{franco} R. Eager, S. Franco and K. Schaeffer, arXiv:1107.1244 [hep-th]; S. Franco,
D. Galloni and Y.-H. He, arXiv:1203.6067 [hep-th].

\bibitem{gn1} A. Gorsky and N. Nekrasov,
  Nucl. Phys. B {\bf 436}, 582 (1995) [hep-th/9401017].

\bibitem{gn2} A. Gorsky and N. Nekrasov,
  Nucl. Phys. B {\bf 414}, 213 (1994) [hep-th/9304047].

\end{thebibliography}
\end{document}